\documentclass[twocolumn,secnumarabic,amssymb, nobibnotes, aps, prb, superscriptaddress]{revtex4-2}
\usepackage{graphicx}%
\usepackage{float}
\usepackage[T1]{fontenc}
\usepackage{color} %white, black, red, green, blue, cyan, magenta, yellow
\usepackage[colorlinks]{hyperref}
\usepackage{amsmath}
\usepackage{amssymb}
\usepackage{bm}
\usepackage[normalem]{ulem}

\begin{document}

%\title{Probing the one-dimensional subband structure of ultrathin core-shell nanowires %in the quantum limit by resonant inelastic light scattering}	

\title{Ultrafast pseudospin quantum beats in multilayer WSe$_2$ and MoSe$_2$}

\author{Simon Raiber}
\affiliation{Institut f\"ur Experimentelle und Angewandte Physik, Universit\"at Regensburg, D-93040 Regensburg, Germany}
\author{Paulo E. Faria~Junior}
\affiliation{Institut f\"ur Theoretische Physik, Universit\"at Regensburg, D-93040 Regensburg, Germany}
\author{Dennis Falter}
\affiliation{Institut f\"ur Experimentelle und Angewandte Physik, Universit\"at Regensburg, D-93040 Regensburg, Germany}
\author{Simon Feldl}
\affiliation{Institut f\"ur Experimentelle und Angewandte Physik, Universit\"at Regensburg, D-93040 Regensburg, Germany}
\author{Petter Marzena}
\affiliation{Institut f\"ur Experimentelle und Angewandte Physik, Universit\"at Regensburg, D-93040 Regensburg, Germany}
\author{Kenji Watanabe}
\affiliation{Research Center for Functional Materials, National Institute for Materials Science, Tsukuba Ibaraki 305-0044, Japan}
\author{Takashi Taniguchi}
\affiliation{International Center for Materials Nanoarchitectonics, National Institute for Materials Science, Tsukuba Ibaraki 305-0044, Japan}
\author{Jaroslav Fabian}
\affiliation{Institut f\"ur Theoretische Physik, Universit\"at Regensburg, D-93040 Regensburg, Germany}
\author{Christian Sch\"uller}
\affiliation{Institut f\"ur Experimentelle und Angewandte Physik, Universit\"at Regensburg, D-93040 Regensburg, Germany}

\date{\today}

\begin{abstract}
Layered van-der-Waals materials with hexagonal symmetry offer an extra degree of freedom to their electrons, the so called valley index or valley pseudospin. This quantity behaves conceptually like the electron spin and the term valleytronics has been coined. In this context, the group of semiconducting transition-metal dichalcogenides (TMDC) are particularly appealing, due to large spin-orbit interactions and a direct bandgap at the K points of the hexagonal Brillouin zone. In this work, we present investigations of excitonic transitions in mono- and multilayer WSe$_2$ and MoSe$_2$ materials by time-resolved Faraday ellipticity (TRFE) with in-plane magnetic fields, $B_{\parallel}$, of up to 9 T. In monolayer samples, the measured TRFE time traces are almost independent of $B_{\parallel}$, which confirms a close to zero in-plane  exciton $g$ factor $g_\parallel$, consistent with first-principles calculations. In stark contrast, we observe pronounced temporal oscillations in multilayer samples for $B_{\parallel}>0$. Remarkably, the extracted in-plane $g_\parallel$ are very close to reported out-of-plane exciton $g$ factors of the materials, namely $|g_{\parallel 1s}|=3.1\pm 0.2$ and $2.5\pm0.2$ for the 1s A excitons in WSe$_2$ and MoSe$_2$ multilayers, respectively. Our first-principles calculations nicely confirm the presence of a non-zero $g_{\parallel}$ for the multilayer samples. We propose that the oscillatory TRFE signal in the multilayer samples is caused by pseudospin quantum beats of excitons, which is a manifestation of spin- and pseudospin layer locking in the multilayer samples. Our results demonstrate ultrafast pseudospin rotations in the GHz- to THz frequency range, which pave the way towards ultrafast pseudospin manipulation in multilayer TMDC samples.

\end{abstract}
\maketitle

The semiconducting transition-metal dichalcogenides (TMDCs) hold great promise for optoelectronic applications, since they form direct bandgap semiconductors in the monolayer limit. Their optical properties are governed by excitons, i.e., Coulomb-bound electron-hole pairs \cite{Chernikov2014,Wang2018}, even at room temperature, due to extraordinarily large exciton binding energies. For high-quality encapsulated MoSe$_2$ monolayers, superior optical quality with exciton linewidths approaching the lifetime limit has been demonstrated \cite{Scuri2018,Back2018}. Starting from bilayers, the bandgap becomes indirect. Nevertheless, going from a single layer to multilayers, the direct interband transitions at the K points of the Brillouin zone still dominate the optical absorption \cite{Ye2015}. Another superior quality of monolayer material is the strong spin-orbit coupling in combination with inversion asymmetry, which lead to large valley-selective spin-orbit splittings of the band edges, culminating in the so called spin-valley locking. This peculiarity is appreciated by the introduction of a pseudospin index, which conceptually behaves like the electron spin, and is connected to the occupation of the two non-equivalent K$^+$ and K$^-$ valleys of the first Brillouin zone. Interestingly, the spin-valley locking of a single layer transforms into a spin- or pseudospin-{\em layer} locking for multilayers \cite{Jones2014}. For TMDC bilayers it has even been suggested that the spin-layer locking can be exploited for the design of spin quantum gates \cite{Gong2013}.

\begin{figure*}
	\includegraphics[width= 1.0\textwidth]{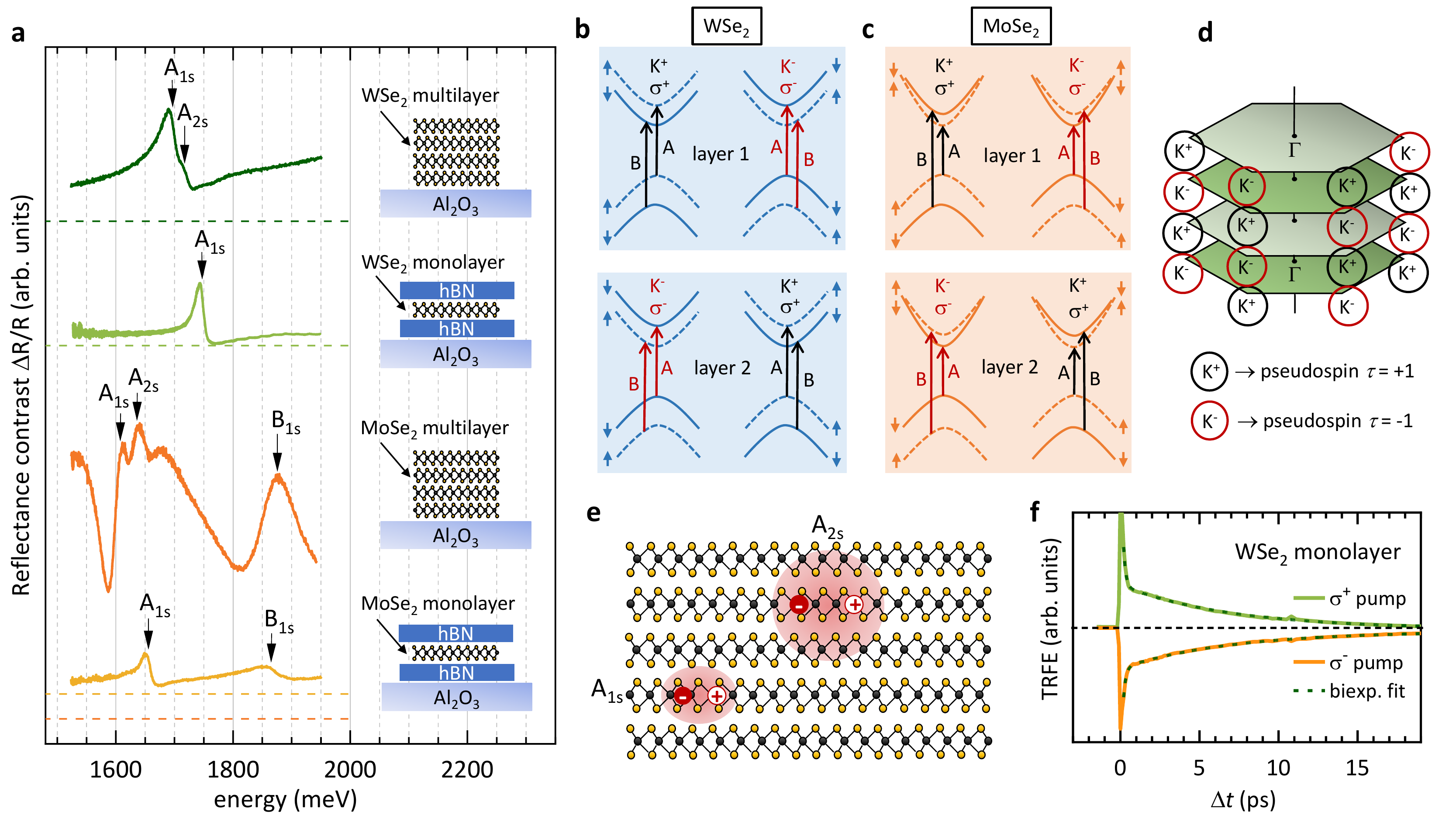}
	\caption{{\bf Reflectance-contrast and TRFE experiments, intralayer transitions, and pseudospin-layer locking.} (a) White-light reflectance-contrast experiments of the investigated samples: MoSe$_2$ and WSe$_2$ monolayers, encapsulated in hBN, and multilayer samples of both materials. The corresponding zero lines are given as dashed lines of the same color. All samples are prepared on transparent sapphire substrates. The substrate temperature in all measurements was $T\sim 20$ K, except for the MoSe$_2$ multilayer, where it was $<10$ K. Excitonic transitions, as derived from a transfer-matrix-model fit, are indicated by small vertical arrows. (b) Schematic picture of momentum- and spin-allowed transitions in an H-type WSe$_2$ bilayer. (c) Same as (b) but for an MoSe$_2$ bilayer. (d) Schematic picture of the layer Brillouin zones in a four-layer structure. Due to the 180$^\circ$ rotation between neighboring layers in an H-type structure, K$^+$ and K$^-$ valleys are alternating. An interlayer pseudospin $\tau=+1$ is connected to the K$^+$ valleys of the individual layers (marked by black circles), while $\tau=-1$ corresponds to the K$^-$ valleys (indicated by red circles). (e) Sketch of intralayer excitons A$_{\rm 1s}$ and A$_{\rm 2s}$. (f) TRFE traces of the encapsulated WSe$_2$ monolayer for both pump helicities, measured in resonance with the A$_{\rm 1s}$ exciton. The dashed lines are biexponential fits to the data.   }
	\label{Fig1}
\end{figure*}

Interlayer excitons (IX), where electron and hole reside in adjacent layers, were first detected in heterobilayers \cite{Rivera2015}. There, the characteristics of IX depend crucially on the material combination \cite{Nagler2017,Nagler2019,Kunstmann2018}. Recently, even valley-polarized currents of IX in heterobilayers have been demonstrated \cite{Unuchek2019}. While in heterobilayers the oscillator strength of IX is weak, the situation can be different for homobilayers or multilayers \cite{Deilmann2018}. In MoS$_2$ bilayers, strong absorption by IX up to room temperature was reported \cite{Calman2018,Slobodeniuk2019,Gerber2019,Niehues2019,Carrascoso2019,Leisgang2020}. 
%It was theoretically suggested that the strong oscillator strength of IX in homobilayers stems from hybridization with the intralayer B exciton \cite{Deilmann2018,Gerber2019}, while only weak mixing of the IX with the intralayer A exciton was found \cite{Leisgang2020}. 
In MoSe$_2$, the situation is similar to MoS$_2$, though the oscillator strength of the IX is  smaller \cite{Horng2018}.
%because of the larger spin-orbit splitting in the valence band, and, hence, larger energetic separation to the B intralayer exciton \cite{Horng2018}. 
Nevertheless, IX have been reported in H-stacked MoTe$_2$ \cite{Arora2017} and MoSe$_2$ \cite{Horng2018,Arora2018} multilayers. %at energies above \cite{Arora2017} and in between \cite{Horng2018} the intralayer 1s and 2s A excitons, A$_{\rm 1s}$ and A$_{\rm 2s}$. %Remarkably, when going from a single layer to a multilayer with $N$ layers, the relation between the oscillator strengths of IX and intralayer A excitons doubles in favor of the IX, since the IX oscillator strength goes with $2(N-1)$ \cite{Horng2018}. 
In contrast to Mo-based multilayers, the momentum-space direct IX in W-based materials has so far not been observed.
%is expected at an energy slightly below the intralayer A$_{\rm 1s}$ exciton. Furthermore, its calculated oscillator strength is smaller than in Mo-based materials \cite{Arora2018}, and it has so far not been identified experimentally. 
It should be noted that for WSe$_2$ homobilayers, IX due to momentum-{\em indirect} transitions below the optical bandgap were reported \cite{Wang2017,Lindlau2018}. 

%While for MoSe$_2$-WSe$_2$ H-type heterobilayers the out-of-plane $g$ factor, $g_\perp$, of the IX is much larger in magnitude than that of the intralayer excitons \cite{Nagler2017g,Wang2020}, in MoSe$_2$ multilayers, the magnitude of $g_\perp$ is similar for both exciton species, A$_{\rm 1s}$ and IX, though its sign is opposite \cite{Arora2018}. 
While monolayer TMDCs have been quite intensely investigated in out-of-plane magnetic fields, investigations on multilayer samples are quite rare. The out-of-plane $g$ factor, $g_\perp$, of the intralayer A excitons is in MoSe$_2$ and WSe$_2$ multilayers smaller in magnitude than in single layers \cite{Arora2018,Arora2018a}. So far, there are, however, no experimental investigations on the {\em in-plane} $g$ factor, $g_\parallel$, in TMDC multilayers available. In-plane magnetic fields, $B_\parallel$, have been applied to TMDC single layers for the brightening of dark excitonic states via mixing of the spin levels by the in-plane field \cite{Zhang2017,Molas2019,Lu2019,Robert2020}. In this work we present time-resolved Faraday ellipticity (TRFE) experiments on WSe$_2$ and MoSe$_2$ mono- and multilayers in {\em in-plane} magnetic fields. While we do not observe a significant influence of in-plane fields of up to 9 T in experiments on monolayers, there is a dramatic effect in multilayer samples: Pronounced temporal oscillations are observed in the TRFE time traces for $B_\parallel >0$. Remarkably, the derived in-plane exciton $g$ factors, $|g_\parallel|$, are close to reported $|g_\perp|$ values of the same materials \cite{Arora2018}. 
%Since there are no oscillations observed for monolayers, we conclude that the observed oscillations are due to ultrafast pseudospin rotations of intralayer excitons, i.e., coherent quantum beats of pseudospin-locked excitons in neighboring layers. 

We start the discussion with reflectance-contrast (RC) experiments of the investigated samples, in order to characterize the excitonic transitions in the materials. Simplified schematic drawings of the first two A excitons, A$_{\rm 1s}$ and A$_{\rm 2s}$, in a multilayer sample are plotted in Fig.~\ref{Fig1}e. Figure \ref{Fig1}a shows an overview of RC spectra of the four samples, investigated in the main body of the manuscript. Excitonic transitions are marked by small vertical arrows as derived from fitting the RC spectra with a transfer-matrix model, assuming complex Lorentz oscillators for the excitonic transitions (see supplementary information). The schematic drawings in the inset of Fig.~\ref{Fig1}a depict the samples, which are MoSe$_2$ and WSe$_2$ mono- and multilayers (for more details, see the methods section). The monolayer samples are encapsulated in hBN to protect them from environmental influences and to provide a homogeneous dielectric environment. The WSe$_2$ multilayer constists of about 14 layers, the MoSe$_2$ multilayer is much thicker, counting about 84 layers, as determined by atomic-force microscopy. In both monolayer samples, the intralayer A$_{\rm 1s}$ excitons show up as distinct and sharp features in the RC spectra in Fig.~\ref{Fig1}a. In the MoSe$_2$ monolayer also the B$_{\rm 1s}$ exciton can be detected, while for the WSe$_2$ monolayer it is outside the displayed energy range. For clarity, the transitions at the K points of the first Brillouin zone, which lead to the excitonic A and B resonances are sketched for both materials in Figs.~\ref{Fig1}b and \ref{Fig1}c (see layer 1, only, for the monolayer case). Interlayer transitions are omitted in the schematic pictures, since they play no role in our experiments. Going from the monolayer to multilayers, the intralayer excitonic resonances show a redshift, and the energetic separation between A$_{\rm 1s}$ and A$_{\rm 2s}$ decreases dramatically because of the stronger dielectric screening \cite{Arora2017,Arora2018}. In agreement with published results \cite{Arora2018}, we observe in Fig.~\ref{Fig1}a in the WSe$_2$ multilayer two features, which can be attributed to the A$_{\rm 1s}$ and A$_{\rm 2s}$ intralayer excitons. This assignment is supported by our excitonic calculations (see methods section): From the derived effective masses for electron and hole we calculate an energetic separation of the A$_{\rm 1s}$ and A$_{\rm 2s}$ intralayer excitons of $\sim 20.8$ meV, which is very close to the experimental value of $\sim 19.2$ meV. 
%As noted above, for WSe$_2$ multilayer the interlayer exciton IX should be weak and is expected at an energy slightly below A$_{\rm 1s}$ \cite{Arora2018,Deilmann2018}. The transitions corrsponding to IX in a WSe$_2$ bilayer are indicated for illustration in Fig.~\ref{Fig1}b by dashed arrows \cite{Gong2013}. Please note that these transitions are indirect in space but direct in momentum space. It would be similar for multilayers.
In agreement with the reports in Ref.~\cite{Arora2018}, we also do not find a feature, related to the IX in WSe$_2$ multilayer in our RC experiments in Fig.~\ref{Fig1}a. 
Also for the MoSe$_2$ multilayer we do not observe a spectral feature, related to the IX.
% There, a distinct feature, related to the IX can be identified by comparison to Refs.~\cite{Arora2018,Horng2018}, where IX has been identified by its positive out-of-plane $g$ factor \cite{Arora2018} and by comparison to advanced calculations, employing the Dirac-Bloch equations \cite{Horng2018}. In Fig.~\ref{Fig1}c, the corresponding transitions are indicated by dashed arrows. Our results in Fig.~\ref{Fig1}a, i.e., the spectral weight and energetic separation of the IX to the intralayer A$_{\rm 1s}$ exciton are in good agreement with the results of Refs.~\cite{Arora2018,Horng2018}. On the other hand, in contrast to Refs.~\cite{Arora2018,Horng2018}, where no A$_{\rm 2s}$ exciton was detected, but 
Similar to Ref.~\cite{Arora2015}, we find spectral features related to the A$_{\rm 1s}$ and A$_{\rm 2s}$ intralayer excitons in the MoSe$_2$ multilayer. Again, this assignment in Fig.~\ref{Fig1}a is corroborated by our computed energy separation of A$_{\rm 1s}$ and A$_{\rm 2s}$ excitons of $\sim 31.8$ meV, which is close to the experimental value of $\sim 28.6$ meV.

Figure \ref{Fig1}d is a sketch of the individual first Brillouin zones of an H-type four-layer structure. In H-type structure, subsequent layers are rotated by 180$^\circ$. Therefore, in momentum space, K$^+$ and K$^-$ valleys of the individual layers are alternating, which is called spin-layer locking \cite{Gong2013,Jones2014}. A pseudospin quantum number $\tau=+1$ $(-1)$ can be attributed to the K$^+$ (K$^-$) valley, leading to a pseudospin-layer locking.

Figure \ref{Fig1}f shows typical TRFE time traces, recorded on the WSe$_2$ monolayer at zero magnetic field under resonant excitation of the A$_{\rm 1s}$ exciton. All experiments presented in this manuscript are in the excitonic regime, i.e., the exciton densities are below the Mott density (see methods section). The light green line shows a trace with $\sigma^+$-polarized pump pulses, which create a K$^+$ valley polarization at time $\Delta t =0$. The orange line is an analogous measurement but with $\sigma^-$ pump pulses, i.e., a K$^-$ valley polarization is initialized. The dashed lines represent biexponential fits to the data. Both measurement curves can be nicely fitted by a biexponential decay with a short time constant of $\tau_r\sim 0.15$ ps and a longer decay time of $\tau_v\sim 7.0$ ps. There are a couple of different processes, which can contribute to the fast decay at short times. Among them is the direct radiative decay of excitons, which are created inside the light cone, and which directly decay radiatively before any scattering event can take place. Our measured $\tau_r$ of $\sim 0.15$ ps is in very good agreement with previous measurements of the radiative lifetime of excitons in WSe$_2$ monolayers \cite{Poellmann2015}. Therefore, it is likely that the fast initial decay of the TRFE signal is influenced by direct radiative recombination of part of the exciton population, created inside the light cone. A significant part of the excitonic population is, however, scattered out of the light cone, e.g., by phonons, and contributes to the valley polarization over a longer time period. The main mechanism leading to valley relaxation in WSe$_2$ monolayers, is the long-range exchange mechanism between electron and hole \cite{Zhu2014,Glazov2014,Yu2014}. The valley-polarization decay time of $\tau_v\sim 7.0$ ps, extracted from the TRFE traces of the hBN encapsulated WSe$_2$ monolayer in Fig.~\ref{Fig1}f, is in very good agreement with the reported decay time of 6.0 ps, measured on a bare WSe$_2$ monolayer on a SiO$_2$ substrate in Ref.~\cite{Zhu2014}.

\begin{figure*}
	\includegraphics[width= 1.0\textwidth]{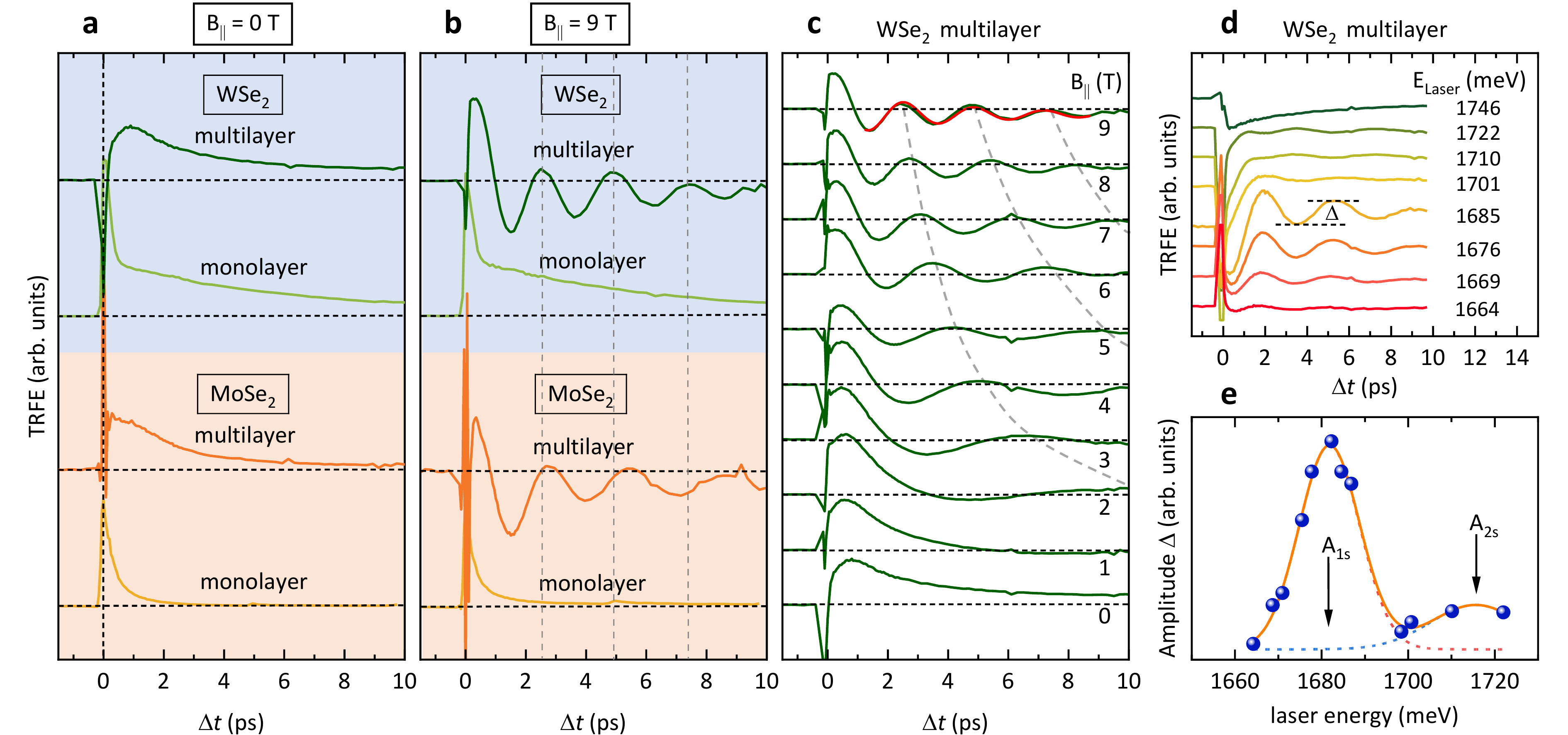}
	\caption{{\bf TRFE experiments in in-plane magnetic fields. |} Comparison of TRFE traces of all samples at (a) $B_\parallel =0$, and, (b) $B_\parallel  =9$ T, excited at the A$_{\rm 1s}$ excitonic resonances. In the multilayer samples, strong temporal oscillations are observed in the time traces at $B_\parallel =9$ T, in contrast to the monolayer samples, which show no oscillations. (c) TRFE traces of the WSe$_2$ multilayer for different in-plane magnetic fields. The red solid line represents an exponentially-damped cosine fit to the data. The dashed gray lines are guides to the eye. (d) TRFE measurements of the WSe$_2$ multilayer at fixed in-plane field $B_\parallel = 6$ T for different center energies $E_{\rm Laser}$ of the laser pulses, as given in the figure. A clear resonance behavior of the signal can be observed. The signal amplitude $\Delta$, as determined for all curves, is indicated. (e) Plot of the extracted signal amplitudes $\Delta$, as indicated in (d), versus central laser energy. The dashed lines represent Gaussian fits, while the solid orange line is the sum of both fit curves. Resonances with the A$_{\rm 1s}$ and A$_{\rm 2s}$ excitons are indicated by arrows. For all measurements the temperature was $T\sim 5$ K.}
	\label{Fig2}
\end{figure*}

We now move on to the central point of the investigations in this manuscript: experiments in in-plane magnetic fields, $B_\parallel$. Figure \ref{Fig2}a shows a comparison of TRFE traces of all four investigated samples at $B_\parallel =0$, where the laser was tuned in resonance with the A$_{\rm 1s}$ excitonic resonances in the respective materials, as marked by arrows in the RC measurements in Fig.~\ref{Fig1}a. The trace of the WSe$_2$ monolayer is the same as shown in Fig.~\ref{Fig1}f ($\sigma^+$ pump). Comparing the two monolayer samples in Fig.~\ref{Fig2}a, one can recognize the much faster valley depolarization in MoSe$_2$. The measured decay time is here $\sim 1$ ps, as compared to $\sim 7$ ps for the WSe$_2$ monolayer (see discussion above). The much faster valley depolarization in the MoSe$_2$ monolayer is reminiscent of a close to zero valley polarization, measured in cw polarized photoluminescence on this material \cite{Tornatzky2018}. Surprisingly, while the valley depolarization time is comparable for the WSe$_2$ monolayer and multilayer, it is much longer in the MoSe$_2$ multilayer, as compared to the monolayer. This may be related to the fact that in MoSe$_2$ monolayers the lowest energy state is a bright state, which is different in all other samples, however, we note here that this is not the focus of this work. In Fig.~\ref{Fig2}b, the same measurements are shown, now for an in-plane field of $B_\parallel = 9$ T. While the TRFE time traces for the monolayer samples are essentially unchanged when compared to $B_\parallel = 0$, they are drastically different for the multilayer samples. Strong and pronounced oscillations can be observed. The oscillation period of the MoSe$_2$ multilayer is slightly longer than for the WSe$_2$ multilayer. As a guide to the eye, vertical dashed lines are plotted in Fig.~\ref{Fig2}b, which mark the maxima of the oscillations of the WSe$_2$ multilayer. Figure \ref{Fig2}c shows a full data set for the WSe$_2$ multilayer from $B_\parallel = 0$ T to 9 T. A full data set of the MoSe$_2$ multilayer is plotted in the supplementary Fig.~S2. The gray dashed lines in Fig.~\ref{Fig2}c are guides to the eye and mark the oscillation maxima, which correspond to the same oscillation period. To test the resonance behavior of the TRFE measurements, we plot in Fig.~\ref{Fig2}d TRFE traces of the WSe$_2$ multilayer at fixed in-plane field of $B_\parallel = 6$ T for different central energies of the laser pulses. The central energies are given in Fig.~\ref{Fig2}d, the spectral widths of the pulses is $\sim 16$ meV. We extract the amplitudes of the oscillations, $\Delta$, as indicated in Fig.~\ref{Fig2}d, and plot them versus central laser energy in Fig.~\ref{Fig2}e. The amplitudes show a clear resonance behavior. The dashed lines in Fig.~\ref{Fig2}e are Gaussian fits, and the solid orange line is the sum of the two Gaussian fit curves. The two maxima can be attributed to resonances with the A$_{\rm 1s}$ and A$_{\rm 2s}$ intralayer excitons (cf.~with the resonance features in the RC experiments in Fig.~\ref{Fig1}a). We note that the A$_{\rm 1s}$ resonance position is shifted by about 16 meV to lower energies in comparison to the white-light RC measurements in Fig.~\ref{Fig1}a, which can be due to bandgap-renormalization effects under pulsed excitation \cite{Steinhoff2017}. For the MoSe$_2$ multilayer we get similar results (not shown). A full dataset of TRFE traces in resonance with the A$_{\rm 2s}$ exciton from 0 to 9T can be found in supplementary Fig.~S3 (same for the MoSe$_2$ multilayer in Fig.~S4). It should be emphasized that we do not observe oscillations, i.e., an excitonic resonance, at energies {\em above} the A$_{\rm 1s}$ and A$_{\rm 2s}$ excitons in the MoSe$_2$ multilayer, in the spectral region, where in Ref.~\cite{Arora2018} an IX was reported in RC measurements. From that we conclude that for our observed temporal oscillations only the intralayer A excitons are relevant. 

\begin{figure}
	\includegraphics[width= 0.5\textwidth]{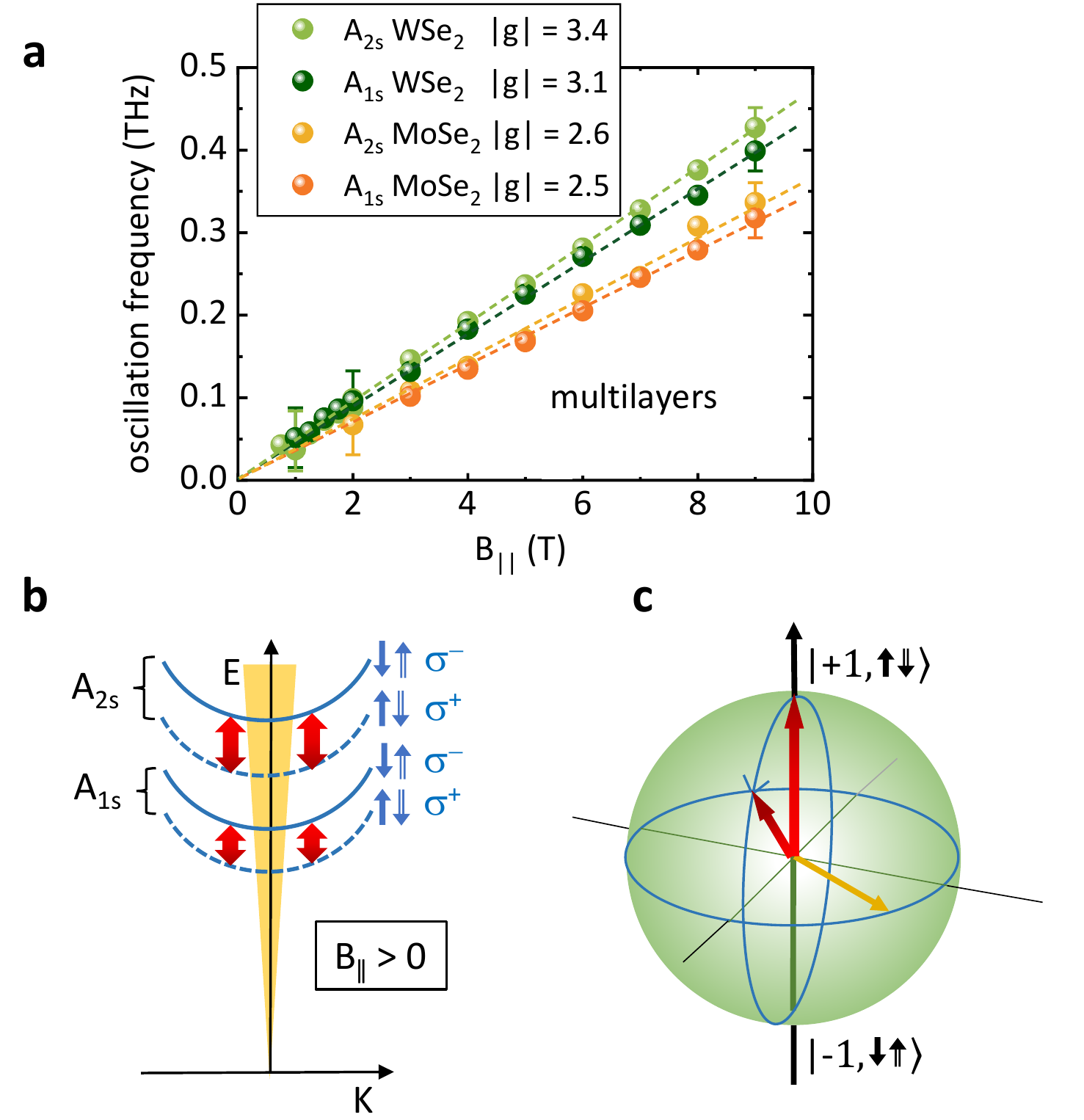}
	\caption{{\bf Extracted in-plane $g$ factors, and proposed mechanism. |} (a) Experimentally detected oscillation frequencies for different intralayer exciton species in the multilayer samples versus in-plane magnetic field. Exemplary error bars for low and high fields are indicated. The extracted absolute values $|g_\parallel |$ are given in the legend. The experimental error margins are about $\pm 0.2$. (b) Energy versus center-of-mass momentum $K$ dispersion of intralayer A excitons at $B_\parallel >0$ in a multilayer. For the excitons, the $z$ component of the spin of the electron is symbolized by a small arrow with a single line, while the hole spin is indicated by an arrow with a double line. Energy splittings of the excitons due to finite $g_\parallel$ are taken into account. The red double arrows should symbolize the coherent momentum-allowed oscillation between adjacent layers. (c) Representation of the pseudospin rotation on a Bloch sphere. The north pole corresponds to $\tau =+1$, while the south pole represents the $\tau = -1$ state. The orange arrow would correspond to a system, excited with linearly-polarized light. }   
	\label{Fig3}
\end{figure}

Clearly, the oscillations in the TRFE traces resemble coherent precession of a magnetic moment about the in-plane magnetic field, as known from, e.g., electron spins in n-doped GaAs bulk \cite{Kikkawa1998}, hole spins in GaAs quantum wells \cite{Gradl2014}, or, localized background charge carriers in MoS$_2$ and WS$_2$ \cite{Yang2015}, among many other examples. We have fitted all experimental curves for $B_\parallel > 0$ with an exponentially-damped cosine function $S(\nu, \tau_v)\propto {\rm exp}(-\Delta t/\tau_v){\rm cos}(2\pi \nu \Delta t)$ for delay times $\Delta t$ well above the fast initial decay of the TRFE signals, as exemplarily shown by the red solid line in Fig.~\ref{Fig2}c for the 9 T trace. An important result is that the oscillations with frequency $\nu$ at $B_\parallel >0$ decay with approximately the same decay time $\tau_v$ as the excitonic signal at $B_\parallel =0$, and no long-lived oscillatory signal is developed. From that we conclude that the oscillations stem from a Larmor precession of the {\em exciton} magnetic moment, and not from the spin of background charge carriers, as observed for localized electrons in MoS$_2$ and WS$_2$ monolayers \cite{Yang2015}. Furthermore, the approximate independence of the decay time $\tau_v$ from $B_\parallel$ shows that $g$ factor fluctuations do not play a role. Otherwise, a $1/B_\parallel$ dependence of  $\tau_v$ would be expected \cite{Yang2015a,Zhukov2020}. Figure \ref{Fig3}a shows a summary of all oscillation frequencies $\nu$, extracted by this procedure, versus $B_\parallel$. Clearly, a linear, Zeeman-like dependence can be recognized. The determined $|g_\parallel|$ are given in the legend of Fig.~\ref{Fig3}a. The experimental error margins for these values are about $\pm 0.2$. It should be noted that with TRFE experiments we can only determine the magnitude of the $g$ factor but not its sign. Very remarkably, for all excitonic resonances, the determined $|g_\parallel |$ are very close to out-of-plane $g$ factors, $|g_\perp|$, of the corresponding materials, reorted in Refs.~\cite{Arora2018,Arora2018a}, which are for WSe$_2$ bulk material $|g_\perp|= 3.2\pm 0.2$ and $3.3\pm 0.6$ for the A$_{\rm 1s}$ and A$_{\rm 2s}$ intralayer excitons, respectively \cite{Arora2018}. For MoSe$_2$ bulk, the reported value for A$_{\rm 1s}$ is $|g_\perp| = 2.7\pm 0.1$ \cite{Arora2018}. Hence, we conclude that $|g_\parallel| \sim |g_\perp|$ for multilayer TMDCs, approaching the bulk limit.

In the following we will discuss our experimental findings further and compare them to first-principles calculations (see methods section).  In table \ref*{table1}, computed spin- and orbital angular momenta for out-of-plane ($\rm S_z^i$, $\rm L_z^i$) as well as in-plane ($\rm S_x^i$, $\rm L_x^i$) directions are given for the monolayer and multilayer samples. The superscript i stands for CB or VB, i.e., for the conduction-band or valence-band states, respectively, which are relevant for the intralayer A excitons in the materials (cf.~Figs.~\ref{Fig1}b and \ref{Fig1}c). We note that the relevant CB states are different for the two materials because of the reverse spin order. The computed $g$ factors for the A excitons, which are determined by $g_{\perp/\parallel} =2(\rm S_{z/x}^{CB}+L_{z/x}^{CB}- S_{z/x}^{VB}-L_{z/x}^{VB})$ \cite{Wozniak2020} are also given. The minus signs in front of the VB angular momenta account for the fact that the angular momentum of a hole is just opposite to the angular momentum of an electron in the VB state. One can see that for the monolayers and WSe$_2$ multilayers, the calculated $g_\perp$ agree well with published experimental values, while for MoSe$_2$ multilayers the computed $|g_\perp|$ is somewhat smaller as compared to the experimental report. The experimental result of $g_\parallel \sim 0$ for the monolayers is confirmed by the calculations, which give exactly $g_\parallel = 0$ (both, spin and orbital angular momenta contributions are zero due to symmetry considerations and verified numerically, cf.~table \ref*{table1}).

For the multilayer samples on the other hand, the calculations do deliver nonzero $g_\parallel$, consistent with our experimental finding, though their magnitudes are smaller than the experimental values, which are close to reported out-of-plane $g$ factors, i.e., $|g_\parallel ({\rm exp.})| \sim |g_\perp (\rm exp.)|$. Interestingly, because of the particular symmetry of the bands (CB $\sim$ $\Gamma_9$ and VB $\sim$ $\Gamma_7$ in the $D_{3h}$ point group of the K valleys), only the valence band shows a nonzero value of $g_\parallel$, while for the conduction band it is strictly zero, i.e.,  the orbital, $\rm L^{CB}_x$, and the spin, $\rm S^{CB}_x$, angular momenta are both zero (cf.~table I). This situation is similar to the zero $g_x$ (Voigt geometry) of the heavy-hole valence band in wurtzite materials with hexagonal symmetry \cite{Rodina2001PRB,Venghaus1977PRB,Tedeschi2019PRB,FariaJunior2019PRB}.
While in the first-principles calculations the interlayer hybridization of electronic bands is fully taken into account, excitonic correlations are not considered. Since we observe the oscillations at excitonic resonances, it is likely that additional hybridization on the excitonic level contribute to the observed $g$ factor. Also the experimental observation that for the multilayers the $|g_\parallel |$ of the 2s excitons are slightly larger than those of the 1s excitons (cf.~Fig.~\ref{Fig3}a) points into this direction: The Bohr radius, i.e., the spatial expansion of the 2s excitons is larger than that of the 1s excitons (see Fig.~\ref{Fig1}e). Therefore, it is likely that hybridization effects may be slightly more important for 2s than for 1s excitons.

\begin{table*}[]
	\begin{tabular}{|c||c c c c||c | c||c c c c||c | c||}
		\hline 
		Material &$\rm S_z^{CB}$ & $\rm L_z^{CB}$ & $\rm S_z^{VB}$ & $\rm L_z^{VB}$ & $g_\perp $ & $g_\perp $ (exp.) & $\rm S_x^{CB}$ & $\rm L_x^{CB}$ & $\rm S_x^{VB}$ & $\rm L_x^{VB}$ & $|g_\parallel |$ & $|g_\parallel|$ (exp.) \\ 
		\hline \hline
		WSe$_2$ monolayer & 0.98 & 2.97 & 1.00 & 5.00 & $-4.10$ & $-4.38 \ldots$$-1.57$ \footnote{Refs.~\cite{Arora2018a,Wang2015,Aivazian2015,Koperski2015,Foerste2020,Robert2017,Chen2019,Srivastava2015,Liu2019}}& 0.00 &0.00 &0.00 &0.00 &0.00 & $\sim 0$ \\ 
		
		WSe$_2$ multilayer  & 0.97 & 2.98 & 1.00 & 4.40 & $-2.89$ & $-3.4 \ldots$$-2.3$ \footnote{Refs.~\cite{Arora2018,Arora2018a,Mitioglu2015}}&0.00 & 0.00 & 0.47 & $\pm 0.07$ & $0.80\ldots 1.08$  & $3.1 \pm 0.2$\\ 
		\hline 
		MoSe$_2$ monolayer & 1.00 & 1.81 & 1.00 & 3.96 & $-4.30$  & $-4.4 \ldots $$-3.8 $ \footnote{Refs.~\cite{Robert2020,Mitioglu2016,Goryca2019,Arora2018a,Ludwig2014,MacNeill2015,Wang2015}} &0.00  &0.00  &0.00  &0.00 &0.00 & $\sim 0$  \\ 
		MoSe$_2$ multilayer & 1.00 & 1.76 & 1.00 & 2.67 & $-1.84$ & $-2.7 $ \footnote{Ref.~\cite{Arora2018}} & 0.00 & 0.00 & 0.74 & $\pm 0.06$ & $1.36\ldots 1.60$  & $2.5 \pm 0.2$\\ 
		\hline 
	\end{tabular}
	\caption{Computed values of out-of-plane and in-plane spin-, S, and orbital, L, angular momenta for the conduction-band (CB) and valence-band (VB) states, which are relevant for the A excitons of the investigated materials. For the first-principles calculations, see the methods section. The corresponding theoretical $g$ factors, $g_\perp$ and $g_\parallel$, for the A excitons are given. For experimental $g_\perp$ of the A$_{\rm 1s}$ exciton, we refer to literature values. The experimental values for $g_\parallel$ from this work are shown in the last column. Since in the experiments we can only determine the magnitude but not the sign, we denote only the magnitude $|g_\parallel|$.}
	\label{table1}
\end{table*}

In Fig.~\ref{Fig3}b, a schematic picture of the excitonic center-of-mass dispersion is shown for the relevant excitonic resonances in the two multilayer materials, namely the A$_{\rm 1s}$ and A$_{\rm 2s}$ resonances. For the excitons, electron and hole spins, $\rm S_z$, are depicted by single-line and double-line arrows, respectively. The energetic splittings, corrsponding to the nonzero $g_\parallel$, are taken into account by dashed and solid lines for the center-of-mass parabolas. The helicities are given next to the spin configurations of the excitons. The bold red double arrows should symbolize the coherent oscillations between the excitonic states, when resonantly excited. Hence, we suggest that the observed oscillations originate from coherent oscillations between excitonic levels with different pseudospins, i.e., pseudospin quantum beats. These can be visualized on a Bloch sphere, as shown in Fig.~\ref{Fig3}c: The north pole corresponds to excitons with pseudospin $\tau = +1$. This means, they occupy the K$^+$ valleys of the individual layers (cf.~Fig.~\ref{Fig1}d). Once they are initialized by a $\sigma^+$ pump pulse, they can coherently oscillate to the south pole, which are excitons with pseudospin $\tau = -1$, i.e., which occupy the K$^-$ valleys of the individual layers. A question, which we can not answer conclusively so far is, if the coherent oscillations are either spin quantum beats of K$^+$ and K$^-$ A excitons solely {\em within} the layers (intralayer oscillations), or {\em between} the layers (interlayer oscillations), or, a mixture of both. The experimental finding that we do not observe oscillations for the monolayers may favor the scenario of interlayer spin quantum beats in the multilayer samples. This is, furthermore, corroborated by the fact that the interlayer component of the oscillations is momentum-allowed, since, in $k$ space K$^+$ and K$^-$ valleys are on top of each other in an H-type structure (cf.~Fig.~\ref{Fig1}d). However, it was previously suggested for WSe$_2$ bilayers that only holes may exhibit coherent oscillations in in-plane magnetic fields \cite{Jones2014}. Presumably, there may be contributions from both, intralayer- and interlayer oscillations. Which part dominates, we can not say so far.

In summary, we have detected ultrafast pseudospin rotations in the GHz to THz frequency range in TMDC multilayers in in-plane magnetic fields via time-resolved Faraday ellipticity. Surprisingly, the magnitudes of the extracted in-plane $g$ factors are close to reported values of out-of-plane $g$ factors of the same materials. This is in stark contrast to monolayer samples, which show no temporal oscillations for nonzero in-plane magnetic field, and which, hence, have an in-plane exciton $g$ factor close to zero. The experimental results are nicely confirmed by first-principles calculations of the $g$ factors. Our study opens the door for manipulation of these pseudospins on ultrafast time scales, making TMDC multilayers an interesting platform for pseudospin operations, possibly putting quantum-gate operations \cite{Jones2014} into reach.

\begin{acknowledgments}
	We gratefully acknowledge valuable discussions with Tobias Korn and would like to thank him for expert help in the initial setup of the experiment. We express our gratitude to Alexey Chernikov for providing the transfer-matrix program, and to Sebastian Bange for expert help with the AFM experiments. Funding by the Deutsche 
	Forschungsgemeinschaft (DFG, German Research Foundation) - Project-ID 
	314695032 - SFB 1277 (subprojects B05, B07 and B11), and projects SCHU1171/8-1 and SCHU1171/10-1 (SPP 2244) is gratefully acknowledged. K.W. and T.T. acknowledge support from JSPS KAKENHI (Grant Numbers 19H05790, 20H00354 and 21H05233).
	
\end{acknowledgments}

\section*{Methods}
\paragraph*{Samples}
All investigated TMDC samples are mechanically exfoliated from bulk source material (purchased from HQ Graphene) using nitto tape, and then transferred onto transparent sapphire substrates by viscoelastic polymethyldisiloxane stamps \cite{Andres}. Large-area MoSe$_2$ and WSe$_2$ monolayers are prepared and encapsulated in hexagonal Boron nitride (hBN) multilayers for protection against environmental influences. In the main body of the manuscript, results from two multilayer samples are presented: A WSe$_2$ multilayer, consisting of 14 layers, and an MoSe$_2$ multilayer with about 84 layers. \\
%For comparison, results from a thicker WSe$_2$ multilayer sample with about 100 layers can be found in the supplementary information (supplementary Fig.~S5).\\

\paragraph*{Optical experiments}
For sample characterization, reflectance-contrast (RC) measurements of all samples are conducted in an optical microscope setup. The samples are mounted by an elastic organic glue on the cold finger of a He-flow cryostat and are kept in vacuum, while the sample holder is cooled down to nominally 5 K. The temperature at the sample position is estimated by the relative intensities of Ruby lines of the sapphire substrate. The substrate temperature is typically between about $T=10$ K and 30 K. For the RC measurements, a white-light source is used, which is focused by a 60x microscope objective to a spot with diameter of about 10 $\mu$m. Reference spectra are recorded at positions next to the TMDC sample. Evaluation of the RC spectra, using a transfer-matrix model, can be found in the supplementary information (supplementary Fig.~S5)

A schematic picture of the experimental setup, used for TRFE experiments, is shown in supplemental Fig.~S1. For TRFE experiments, a mode-locked Ti:Sapphire laser is used, which produces laser pulses with a temporal length of about 80 fs at a repetition rate of 80 MHz. The laser beam is divided into two pulse trains by a beam splitter. The time delay, $\Delta t$, between pump and probe pulses is adjusted by a retroreflector, which is mounted on a linear stepper stage. Both beams are focused by a plano convex lens onto the sample surface, where they overlap. The laser spot diameter at the sample position is about 50 $\mu$m. The sample is mounted in an optical cryostat with superconducting magnet coils (split-coil cryostat) at a temperature of about $T=5$ K, which is maintained by a constant flow of cold He gas. By measuring the laser pulse length before and after the magnet cryostat, we estimate the pulse length at the sample position to be about 130 fs. The pump pulses are circularly polarized and the laser wavelength is tuned to excitonic absorption lines to create a valley polarization in the sample. The temporal dynamics of the valley polarization is then measured by detecting the ellipticity of the linearly-polarized probe pulses after transmission of the sample. For measurement of the ellipticity, a combination of a Wollaston prism, quarter-wave plate and two balanced photo diodes is used. The pump beam is mechanically chopped at a frequency of about 1.6 kHz, and for detection of the photodiode difference signal, lockin technique is used.\\

\paragraph*{Exciton densities}
To get a most accurate estimate of the exciton densities in the experiments, we measure the power of the transmitted pump laser beam for the two cases, when (i) the pump beam is focused on the sample, and (ii) focused next to the sample on the sapphire substrate. The difference in power is the upper limit of the power absorbed by the sample, since with this approach we neglect the difference in reflectivity of the sapphire substrate versus sapphire substrate with TMDC sample. We then assume that the exciton density $n$ is equal to the density of absorbed photons $n_{\rm photons}$, which is related to the absorbed power $P_{\rm abs}$ by $P_{\rm abs}=n_{\rm photons}E_{\rm Laser}fr^2\pi$. $E_{\rm Laser}$ is the energy of the laser photons, $f$ the repetition rate (80 MHz) of the laser, and $r=25$ $\mu$m the laser-spot radius on the sample.

With this procedure, we get for the WSe$_2$ monolayer an initial exciton density of $n\sim 1.3\times 10^{12}$ cm$^{-2}$ and for the MoSe$_2$ monolayer $n\sim 1.9\times 10^{12}$ cm$^{-2}$, when in both cases the A$_{\rm 1s}$ exciton is excited resonantly. Both values are well below the Mott density \cite{Dendzik2020,Steinhoff2017,Siday2022}. For the multilayer samples we devide the total exciton density by the number of layers to get an estimate of the density per layer. We get for the WSe$_2$ multilayer (14 layers) $n\sim 3.7\times 10^{11}$ cm$^{-2}$/layer when exciting the A$_{\rm 1s}$ exciton resonantly, and, $n\sim 2.1\times 10^{11}$ cm$^{-2}$/layer for resonant excitation at the A$_{\rm 2s}$ exciton. For the MoSe$_2$ multilayer (80 layers), we have $n\sim 1.0\times 10^{12}$ cm$^{-2}$/layer for the A$_{\rm 1s}$ exciton resonance, and $n\sim 1.4\times 10^{12}$ cm$^{-2}$/layer for the A$_{\rm 2s}$ exciton resonance.\\

\paragraph*{Theoretical modeling}
The first-principles calculations are performed within the density functional 
theory (DFT) using the full-potential all-electron code WIEN2k \cite{wien2k}. 
We use the Perdew-Burke-Ernzerhof (PBE) exchange-correlation functional \cite{Perdew1996PRL}, 
a core–valence separation energy of $-6$ Ry, atomic spheres with orbital quantum 
numbers up to 10 and the plane-wave cutoff multiplied by the smallest atomic radii 
is set to 9. For the inclusion of spin-orbit coupling, core electrons are considered 
fully relativistically whereas valence electrons are treated in a second variational 
step \cite{Singh2006}. We use a Monkhorst-Pack k-grid of 15$\times$15$\times$6 (15$\times$15) 
for the bulk (monolayer). The bulk calculations include van der Waals interactions via the 
D3 correction \cite{Grimme2010JCP}. Self-consistency convergence was achived using the 
criteria of 10$^{-6}$ e for the charge and 10$^{-6}$ Ry for the energy. The bulk lattice 
parameters, taken from Ref.~\cite{Kormanyos2015TDM}, are $a = 3.282 \; \textrm{\AA}$, 
$d = 3.340 \; \textrm{\AA}$ and $c = 12.960 \; \textrm{\AA}$ for WSe$_2$; and 
$a = 3.289 \; \textrm{\AA}$, $d = 3.335 \; \textrm{\AA}$ and $c = 12.927 \; \textrm{\AA}$
for MoSe$_2$. Here, the in-plane lattice parameter, $a$, and the layer thickness, $d$, 
are considered the same for bulk and monolayers. In monolayers, we used a vacuum 
spacing of 16 $\textrm{\AA}$ to avoid interaction among the periodic replicas whereas 
in the bulk case the total size of the unit cell is the lattice parameter $c$. The 
calculations of the orbital angular momenta $L_x$ and $L_z$ are based on the fully 
converged summation-over-bands approach discussed in Refs.~\cite{Wozniak2020,Deilmann2020PRL,Foerste2020,Xuan2020PRR}.

For the calculations of the bulk intralayer excitons we used the effective Bethe-Salpeter 
equation \cite{Rohlfing200PRB,Zollner2019PRB}. The energy band dispersion near the K valley is treated as 
$E(k_x,k_y,k_z) = \frac{\hbar^2}{2m^*}\left(k^2_x+k^2_y\right) + f(k_z)$, with $m^*$ 
being the in-plane effective mass and $f(k_z)$ models the dispersion from along 
the $-$H $-$ K $-$ H direction of the bulk first Brillouin zone. The DFT calculated 
in-plane effective masses for WSe$_2$ are $m_\text{CB} = 0.29m_0$ and $m_\text{VB} = 0.36m_0$, 
and for MoSe$_2$, $m_\text{CB} = 0.90m_0$ and $m_\text{VB} = 0.61m_0$. For the function $f(k_z)$,  
we take the numerical values directly from the DFT calculations. The electron-hole 
interaction is mediated by the anisotropic Coulomb potential, with the dielectric 
constants for WSe$_2$ given by $\varepsilon_{xx} = \varepsilon_{yy} = 15.75$ 
and $\varepsilon_{zz} = 7.75$, and for MoSe$_2$, $\varepsilon_{xx} = \varepsilon_{yy} = 17.45$ 
and $\varepsilon_{zz} = 8.3$, taken from Ref.~\cite{Laturia2018npj2D}. Our calculations 
reveal binding energies of 29.9 (9.1) meV for the $A_{1s(2s)}$ exciton in WSe$_2$ and 41.9 
(10.1) meV for the $A_{1s(2s)}$ exciton in MoSe$_2$, respectively. Further details on 
this approach for intralayer excitons in bulk TMDCs can be found in Ref.~\cite{Oliva2022arXiv} 
for bulk WS$_2$.

\section*{Contributions}
S.R., D.F., S.F. and P.M. prepared the samples, performed the experiments and analyzed the data. P.E.F.J. and J.F. performed the first-principles and exciton calculations. K.W. and T.T. supported the high-quality hBN material. C.S. conceived the project, analyzed the data and wrote the manuscript. All authors contributed to the discussion of results and to the finalization of the manuscript.


\begin{thebibliography}{00}
\bibitem{Chernikov2014}Chernikov, A. et al. Exciton binding energy and nonhydrogenic Rydberg
series in monolayer WS$_2$. Phys. Rev. Lett. {\bf 113}, 076802 (2014).

\bibitem{Wang2018}Wang, G. et al. Colloquium: excitons in atomically thin transition metal
dichalcogenides. Rev. Mod. Phys. {\bf 90}, 021001 (2018).	

\bibitem{Scuri2018}Scuri, G. et al. Large excitonic reflectivity of monolayer MoSe$_2$ encapsulated
in hexagonal boron nitride. Phys. Rev. Lett. {\bf 120}, 037402 (2018).

\bibitem{Back2018}Back, P. et al. Realization of an
electrically tunable narrow-bandwidth atomically thin mirror using
monolayer MoSe$_2$2. Phys. Rev. Lett. {\bf 120}, 037401 (2018).

\bibitem{Ye2015}Ye, M. et al. Recent Advancement on the Optical Properties of Two-Dimensional Molybdenum Disulfide (MoS$_2$) Thin Films. Photonics {\bf 2}, 288 (2015).

\bibitem{Jones2014}Jones, A. M. et al. Spin–layer locking effects in optical orientation of
exciton spin in bilayer WSe$_2$. Nat. Physics {\bf 10}, 130 (2014).

\bibitem{Gong2013}Gong, Z. et al. Magnetoelectric effects and valley-controlled spin quantum
gates in transition metal dichalcogenide bilayers. Nat. Commun. {\bf 4},
2053 (2013).



\bibitem{Rivera2015}Rivera, P. et al. Observation of long-lived interlayer excitons in monolayer
MoSe2-WSe2 heterostructures. Nat. Commun. {\bf 6}, 6242 (2015).

\bibitem{Nagler2017}Nagler, P. et al. Interlayer exciton dynamics in a dichalcogenide monolayer
heterostructure. 2D Mater. {\bf 4}, 025112 (2017).

\bibitem{Nagler2019}Nagler, P. et al. Interlayer excitons in transition-metal dichalcogenide
heterobilayers. Phys. Status Solidi B {\bf 256}, 1900308 (2019).

\bibitem{Kunstmann2018}Kunstmann, J. et al. Momentum-space indirect interlayer excitons in transition-metal dichalcogenide van der Waals heterostructures. Nat. Phys. {\bf 14}, 801 (2018).

\bibitem{Unuchek2019}Unuchek, D. et al. Valley-polarized exciton currents in a van der Waals
heterostructure. Nat. Nanotechnol. {\bf 14}, 1104 (2019).

\bibitem{Deilmann2018}Deilmann, T. Tygesen, K. S. Interlayer excitons with large optical
amplitudes in layered van der Waals materials. Nano Lett. {\bf 18}, 2984 (2018).

\bibitem{Calman2018}Calman, E. et al. Indirect excitons in van der Waals heterostructures at room
temperature. Nat. Commun. {\bf 9}, 1895 (2018).

\bibitem{Slobodeniuk2019}Slobodeniuk, A. et al. Fine structure of K-excitons in multilayers of transition metal dichalcogenides. 2D Mater. {\bf 6}, 025026 (2019).

\bibitem{Gerber2019}Gerber, I. C. et al. Interlayer excitons in bilayer MoS$_2$ with strong oscillator
strength up to room temperature. Phys. Rev. B {\bf 99}, 035443 (2019).

\bibitem{Niehues2019}Niehues, I. et al. Interlayer excitons in bilayer MoS$_2$ under uniaxial tensile strain. Nanoscale {bf 11}, 12788 (2019).

\bibitem{Carrascoso2019}Carrascoso, F. et al. Biaxial strain tuning of interlayer excitons in bilayer MoS$_2$. J. Phys. Mater. {\bf 3}, 015003 (2019).

\bibitem{Leisgang2020}Leisgang, A. et al. Giant Stark splitting of an exciton in bilayer Mo$_2$, Nat. Nanotechnol. {\bf 15}, 901 (2020).

\bibitem{Horng2018}Horng, J. et al. Observation of interlayer excitons in MoSe$_2$ single crystals. Phys. Rev. B {\bf 97}, 241404 (2018).

\bibitem{Arora2017}Arora, A. et al. Interlayer excitons in a bulk van der Waals semiconductor.
Nat. Commun. {\bf 8}, 639 (2017).

\bibitem{Arora2018}Arora, A. et al. Valley-contrasting optics of interlayer excitons in Mo-and W-based bulk transition metal dichalcogenides. Nanoscale {\bf 10}, 15571 (2018).

\bibitem{Wang2017}Wang, Z. et al. Electrical tuning of
interlayer exciton gases in WSe$_2$ bilayers. Nano Lett. {\bf 18}, 137 (2017).

\bibitem{Lindlau2018}Lindlau, J. et al. The role of momentum-dark excitons in the elementary
optical response of bilayer WSe$_2$. Nat. Commun. {\bf 9}, 2586 (2018).


\bibitem{Arora2018a}Arora, A. et al. Zeeman spectroscopy of
excitons and hybridization of electronic states in few-layer
WSe$_2$, MoSe$_2$, and MoTe$_2$, 2D Mater. {\bf 6}, 015010 (2019).


%\bibitem{Nagler2017g}Nagler, P. et al. Giant magnetic splitting inducing near-unity valley polarization in van der Waals heterostructures. Nat. Commun. {\bf 8}, 1551 (2017).
%\bibitem{Wang2020}Wang, T. Giant Valley-Zeeman Splitting from Spin-Singlet and Spin-Triplet
%Interlayer Excitons in WSe$_2$/MoSe$_2$ Heterostructure. Nano Lett. {\bf 20}, 694 (2020).

\bibitem{Zhang2017}Zhang, X.-X. et al. Magnetic brightening and control of dark excitons
in monolayer WSe$_2$, Nat. Nanotechnol. {\bf 12}, 883 (2017).

\bibitem{Molas2019}Molas, M. R. et al. Probing and Manipulating Valley Coherence of Dark Excitons in Monolayer WSe$_2$. Phys. Rev. Lett. {\bf 123}, 096803 (2019).

\bibitem{Lu2019}Lu, Z. et al. Magnetic field mixing and splitting of bright and dark excitons in monolayer MoSe$_2$. 2D Materials {\bf 7}, 015017 (2019).

\bibitem{Robert2020}Robert, C. et al. Measurement of the spin-forbidden dark excitons
in MoS$_2$ and MoSe$_2$ monolayers. Nat. Commun. {\bf 11}, 4037 (2020).

\bibitem{Arora2015}Arora, A. et al. Exciton band structure in layered MoSe$_2$: from a
monolayer to the bulk limit. Nanoscale {\bf 7}, 20769 (2015).


\bibitem{Poellmann2015}Poellmann, C. et al. Resonant internal quantum transitions and femtosecond radiative decay of excitons in monolayer WSe$_2$. Nat. Materials {\bf 14}, 889 (2015).

\bibitem{Zhu2014}Zhu, C. R. et al. Exciton valley dynamics probed by Kerr rotation in WSe$_2$ monolayers. Phys. Rev. B {\bf 90}, 161302 (2014).

\bibitem{Glazov2014}Glazov, M. M. et al. Exciton fine structure and spin decoherence in monolayers of transition metal dichalcogenides. Phys. Rev. B {\bf 89}, 201302 (2014).

\bibitem{Yu2014}Yu, T. and Wu, M. W.  Valley depolarization due to intervalley and intravalley electron-hole exchange interactions in monolayer 
MoS$_2$. Phys. Rev. B {\bf 89}, 205303 (2014).

\bibitem{Tornatzky2018}Tornatzky, H. et al. Resonance Profiles of Valley Polarization in Single-Layer MoS$_2$ and MoSe$_2$. Phys. Rev. Lett. {\bf 121}, 167401 (2018).

\bibitem{Steinhoff2017}Steinhoff, A. et al. Exciton fission in monolayer transition metal dichalcogenide semiconductors. Nat. Commun. {\bf 8}, 1166 (2017).

\bibitem{Kikkawa1998}Kikkawa, J.M. and Awschalom, D. Resonant Spin Amplification in n-Type GaAs. Phys. Rev. Lett. {\bf 80}, 4313 (1998).

\bibitem{Gradl2014}Gradl, C. et al. Hole-spin dynamics and hole 
g-factor anisotropy in coupled quantum well systems. Phys. Rev. X {\bf 90}, 165439 (2014).

\bibitem{Yang2015}Yang, L. et al. Long-lived nanosecond spin relaxation and spin coherence of electrons in monolayer MoS$_2$ and WS$_2$. Nat. Phys. {\bf 11}, 830 (2015).

\bibitem{Yang2015a}Yang, L. et al. Spin coherence and dephasing of localized electrons in monolayer MoS$_2$. Nano Lett. {\bf 15}, 8250 (2015).

\bibitem{Zhukov2020}Zhukov, E. A. et al. Renormalization of the electron g factor in the degenerate two-dimensional electron gas of ZnSe- and CdTe-based quantum wells. Phys. Rev. B {\bf 102}, 125306 (2020).

\bibitem{Wang2015}Wang, G. et al. Magneto-optics in transition metal diselenide monolayers. 2D Materials {\bf 2}, 034002 (2015).

\bibitem{Aivazian2015}Aivazian, G. et al. Magnetic
control of valley pseudospin in monolayer WSe$_2$. Nat. Phys. {\bf 11},
148 (2015).

\bibitem{Koperski2015}Koperski, M. et al. Single photon emitters in exfoliated WSe$_2$ structures. Nat. Nanotechnol. {\bf 10}, 503 (2015).

\bibitem{Robert2017}Robert, C. et al. Fine structure and lifetime of dark excitons in transition metal dichalcogenide monolayers. Phys. Rev. B {\bf 96}, 155423 (2017).

\bibitem{Chen2019}Chen, S.-Y. et al. Luminescent emission of excited Rydberg excitons from monolayer WSe$_2$, Nano Lett. {\bf 19}, 2464 (2019).

\bibitem{Srivastava2015}Srivastava, A. et al. Valley Zeeman effect in elementary
optical excitations of monolayer WSe$_2$, Nat. Phys. {\bf 11}, 141
(2015).

\bibitem{Liu2019}Liu, E. et al. Magnetophotoluminescence of exciton Rydberg
states in monolayer WSe$_2$, Phys. Rev. B {\bf 99}, 205420 (2019).

\bibitem{Foerste2020}F\"orste, J. et al. Exciton g-factors in monolayer and bilayer WSe$_2$ from experiment and theory. Nat. Commun. {\bf 11}, 4539 (2020).

\bibitem{Mitioglu2015}Mitioglu, A. A. et al. Optical Investigation of Monolayer and Bulk Tungsten Diselenide (WSe$_2$) in High Magnetic Fields. Nano Lett. {\bf 15}, 4387 (2015).

\bibitem{Mitioglu2016}Mitioglu, A. A. et al. Magnetoexcitons in large area CVD-grown monolayer MoS$_2$ and MoSe$_2$ on sapphire. Phys. Rev. B {\bf 93}, 165412 (2016).

\bibitem{Goryca2019}Goryca, M. et al. Revealing exciton masses and dielectric properties of monolayer semiconductors with high magnetic fields. Nat.
Commun. {\bf 10}, 4172 (2019).

\bibitem{Ludwig2014}Li, Y. et al. Valley Splitting and
Polarization by the Zeeman Effect in Monolayer MoSe$_2$, Phys.
Rev. Lett. {\bf 113}, 266804 (2014).

\bibitem{MacNeill2015}MacNeill, D. et al. Breaking
of Valley Degeneracy by Magnetic Field in Monolayer MoSe$_2$,
Phys. Rev. Lett. {\bf 114}, 037401 (2015).

\bibitem{Wozniak2020}Wozniak, T. et al. Exciton g factors of van der Waals heterostructures from first-principles calculations. Phys. Rev. {\bf 101}, 235408 (2020).

\bibitem{Venghaus1977PRB}Venghaus, H. et al. Magnetoluminescence and magnetoreflectance of the $A$ exciton of CdS and CdSe. Phys. Rev. B {\bf 16}, 4419 (1977).


\bibitem{Rodina2001PRB}Rodina, A. V. et al. Free excitons in wurtzite GaN. Phys. Rev. B {\bf 64}, 115204 (2001).


\bibitem{Tedeschi2019PRB}Tedeschi, D. et al. Unusual spin properties of InP wurtzite nanowires revealed by Zeeman splitting spectroscopy. Phys. Rev. B {\bf 99}, 161204 (2019).

\bibitem{FariaJunior2019PRB}Faria~Junior, P. E. et al. Common nonlinear features and spin-orbit coupling effects in the Zeeman splitting of novel wurtzite materials. Phys. Rev. B {\bf 99}, 195205 (2019).

\bibitem{Andres}Castellanos-Gomez, A. et al. Deterministic transfer of two-dimensional materials by all-dry
viscoelastic stamping. 2D Materials {\bf 1}, 011002 (2014).



%\bibitem{Snelling}Snelling, M. J. et al. Exciton, heavy-hole, and electron g factors in type-I GaAs/AlGaAs quantum wells. Phys. Rev. B {\bf 45}, 3922(R) (1992).

%\bibitem{Nye}Nye, J. F. Physical Properties Of Crystals: Their Representation by Tensors and Matrices, Oxford university press (1957).


\bibitem{Dendzik2020}Dendzik, M. et al. Observation of an Excitonic Mott Transition through Ultrafast Core-cum-Conduction Photoemission Spectroscopy. Phys. Rev. Lett. {\bf 125}, 096401 (2020).

\bibitem{Siday2022}Siday, T. et al. Ultrafast Nanoscopy of High-Density Exciton Phases in WSe$_2$. Nano Lett. doi.org/10.1021/acs.nanolett.1c04741 (2022).

\bibitem{wien2k}Blaha, P. et al. An APW+ lo program for calculating the properties of solids. The Journal of chemical physics {\bf 152}, 074101 (2020).

\bibitem{Perdew1996PRL}Perdew, J. P. et al. Generalized Gradient Approximation Made Simple. Phys. Rev. Lett. {\bf 77}, 3865 (1996).

\bibitem{Singh2006}Singh, D. J. and Nordstrom, L. Planewaves, Pseudopotentials, and the LAPW method. Springer Science \& Business Media (2006).

\bibitem{Grimme2010JCP}Grimme, S. et al. A consistent and accurate ab initio parametrization of density functional dispersion correction (DFT-D) for the 94 elements H-Pu. The Journal of chemical physics {\bf 132}, 154104 (2010).

\bibitem{Kormanyos2015TDM}Korm{\'a}nyos, A. et al. k{\textperiodcentered} p theory for two-dimensional transition metal dichalcogenide semiconductors. 2D Materials {\bf 2},
022001 (2015).


\bibitem{Deilmann2020PRL}Deilmann, T. et al. Ab Initio Studies of Exciton $g$ Factors: Monolayer Transition Metal Dichalcogenides in Magnetic Fields. Phys. Rev. Lett. {\bf 124}, 226402 (2020).

\bibitem{Xuan2020PRR}Xuan, F. and Quek, S. Y. Valley Zeeman effect and Landau levels in two-dimensional transition metal dichalcogenides. Phys. Rev. Research {\bf 2}, 033256 (2020).

\bibitem{Rohlfing200PRB}Rohlfing, M. and Louie, S. G. Electron-hole excitations and optical spectra from first principles. Phys. Rev. B {\bf 62}, 4927 (2000).


\bibitem{Zollner2019PRB}Zollner, K. et al. Proximity exchange effects in MoSe$_2$ and WSe$_2$ heterostructures with CrI$_3$: Twist angle, layer, and gate dependence. Phys. Rev. B {\bf 100}, 085128 (2019).

\bibitem{Laturia2018npj2D}Laturia, A. et al. Dielectric properties of hexagonal boron nitride and transition metal dichalcogenides: from monolayer to bulk. npj 2D Materials and Applications {\bf 2}, 1 (2018).


\bibitem{Oliva2022arXiv}Oliva, R. et al. Strong substrate effects in multilayered WS2 revealed by high-pressure optical measurements. arXiv:2202.08551 (2022).


\end{thebibliography}
\end{document}